\documentclass{vgtc}                          

\ifpdf
  \pdfoutput=1\relax                   
  \usepackage{graphicx}                
  \DeclareGraphicsExtensions{.pdf,.png,.jpg,.jpeg} 
\else
  \usepackage{graphicx}                
  \DeclareGraphicsExtensions{.eps}     
\fi%

\graphicspath{{figures/}{pictures/}{images/}{./}} 

\usepackage{microtype}                 
\PassOptionsToPackage{warn}{textcomp}  
\usepackage{textcomp}                  
\usepackage{mathptmx}                  
\usepackage{times}                     
\usepackage{cite}                      
\usepackage{tabu}                      
\usepackage{booktabs}                  
\usepackage{svg}
\usepackage[font=small,skip=0pt]{caption}

\onlineid{0}

\vgtccategory{Research}

\vgtcinsertpkg

\title{How Learners Sketch Data Stories}

\author{Rahul Bhargava\thanks{e-mail: r.bhargava@northeastern.edu}\\ %
        \scriptsize Northeastern University %
\and Dee Williams\thanks{e-mail: williams.dei@northeastern.edu}\\ %
     \scriptsize Northeastern University %
\and Catherine D'Ignazio\thanks{e-mail: dignazio@mit.edu}\\ %
     \parbox{1.4in}{\scriptsize \centering MIT}}

\abstract{Learning data storytelling involves a complex web of skills. Professional and academic educational offerings typically focus on the computational literacies required, but professionals in the field employ many non-technical methods; sketching by hand on paper is a common practice. This paper introduces and classifies a corpus of 101 data sketches produced by participants as part of a guided learning activity in informal and formal settings. We manually code each sketch against 12 metrics related to visual encodings, representations, and story structure. We find evidence for preferential use of positional and shape-based encodings, frequent use of symbolic and textual representations, and a high prevalence of stories comparing subsets of data. These findings contribute to our understanding of how learners sketch with data. This case study can inform tool design for learners, and help create educational programs that introduce novices to sketching practices used by experts.%
} 

\CCScatlist{
  \CCScatTwelve{Human-centered computing~Visualization design and evaluation methods}{Social and professional topics~Informal education}{}{}
}

\begin{document}


\firstsection{Introduction}

\maketitle


The last 20 years has seen massive growth in scholarly work on data literacy, with significant attention paid to introducing this skill to learners in academic \cite{timmermann_critical_2020, klenke_curriculum_2020} and professional settings \cite{qlik_developing_2018}. Much of this cross-field push has focused on technical skill development, such as teaching programming \cite{matei_democratizing_2017, domik_we_2000}, data analysis and visualization tools \cite{lo_learning_2019, batt_learning_2020}, graphic design \cite{chong_aligning_2012, dur_data_2014}, and GIS tools \cite{han_web_2019}. While technical skills are important, focusing on them to the exclusion of other aspects of data literacy doesn’t address some issues novices may have with approaching and understanding data.

We argue there is more work needed on evaluating the non-technological practices often employed by information designers. A practice used consistently by data experts when creating visual stories is sketching their design ideas by hand. Most professionals have notebooks full of ideas for visual depictions of symbols, encodings, and narrative flows \cite{mcnamara_drawing_2021}. But how do learners new to data visualization and storytelling get started with the practice of sketching? What cognitive models and design ideas do they bring to their first sketches of data stories? How can those inform educational activities and tool design for novices in the field?

\begin{figure}[htbp]
  \centering
  \includegraphics[width=9cm]{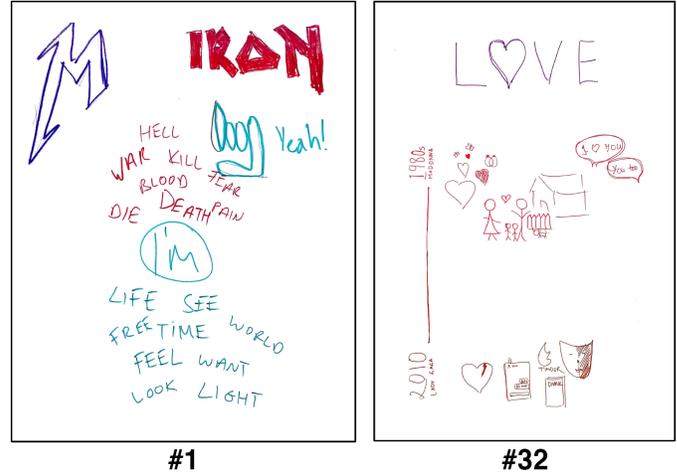}
  \caption{Two sketches from our corpus. \#1 includes textual representations, a comparison story, and positional/color encodings. \#32 includes mostly symbolic representations, a change over time story, and color, positional, and shape encodings.}
  \label{samples}
\end{figure}

To tackle these questions, in this paper we analyze a sample of drawings made by learners as part of a “Sketch a Story” activity. The exercise focuses on helping novices learn how to quickly move from text data, to quantitative analysis, to a visual sketch of a story based on what they see in that data. We introduce a set of metrics for classifying the sketches and code a random sample of 101 hand-drawn sketches produced in these workshops (Figure \ref{samples}). Our findings can inform further work to understand how novices approach sketching data stories, assist tool-designers that want to support learners entering the field, and identify ways to design learning activities.

\section{Related Work}

There is existing work in various domains we can build on in our effort to better understand how learners move into the practice of sketching visual depictions of data. This can be grounded first in existing conceptions of “visual literacy”. Kedra summarizes and defines this as the ability to understand and create meaning via visual stimuli \cite{kedra_what_2018}. Within this paradigm, the sketches produced by participants in this activity exemplify translation (the transformation of text into image) and visual writing (the creation of meaningful images). Though data literacy and visual literacy are not the same, they’re certainly interconnected, and understanding data visualizations requires a combination of visual literacy, numeracy, and statistical understanding \cite{schield_information_2004}.

We can similarly build on a foundational understanding of the role of drawing in learning in general. Dix, \textit{et al} \cite{dix_externalisation_2011} studies this in the field of design, while Tytler, \textit{et al} \cite{tytler_drawing_2020} studies science learning in school. Despite the authors’ different fields, the findings are similar: the act of drawing can aid reasoning and create meaning in a variety of different ways. For examples, the role of externalization is particularly relevant - the creation of physical artifacts to represent mental concepts stimulates interaction and dialog based on a shared perception. Roberts, \textit{et al} formalizes this in their well known “Five Sheet Design Methodology” \cite{roberts_sketching_2016}, which places sketching at the center of a process of exploring and evaluating design alternatives to create interactive information visualizations. Jansson, \textit{et al}’s concept of “design fixation” \cite{jansson_design_1991} also suggests that sketching might be one approach to overcoming the pre-existing sets of ideas that limit conceptualizing a broad set of design alternatives. This line of research connects deeply with the pedagogy and practice put into play in our activity, beyond serving as examples of the impacts the role sketching plays for practitioners as they move from data to story.

From the fields of computer graphics and design we find inspiration in past research on how novices create visualizations. In general the findings related to learner-created visual depictions of data suggest that they tend to rely on familiar chart types \cite{walny_visual_2011}, and sometimes struggle in manipulating data to find answers to the questions they have \cite{bressa_sketching_2019}. This has implications for how activities and platforms for novice visualizations should be built and framed \cite{grammel_how_2010}. In situated visualizations designed for concrete goals, it was harder to get the participants in the novice group to represent things visually in place of using text labels. In addition, work from Wang, \textit{et al} suggests that confidence, or "self-efficacy", in sketching skills is a barrier to entering the practice of sketching effectively and compellingly when visualizing data stories \cite{wang_teaching_2019}.

Perhaps the most connected example of prior work on novices sketching data is from Walny, \textit{et al}, who led participants through a sketching activity and assessed what they produced \cite{walny_exploratory_2015}. They created metrics to assess representations (from numeric to abstract) and hypotheses embedded in a small study of sketches. We extend and reinforce their work, while contributing a larger sample of sketches as another case study to deepen our understanding.

\section{Methodology}

Educators in the field of data visualization and storytelling play a central role in introducing novices to common practices used by experts \cite{kerren_teaching_2008}. Studying and building on the approaches, methods, and mindsets learners bring to the table can help drive appropriate and effective activity design in educational settings. Our corpus is a set of 101 hand-made data sketches created by learners in informal and formal education settings. These workshops all included the Sketch a Story activity described below, were hosted over the last 10 years, and were facilitated by some combination of the authors. We classify and evaluate these sketches via qualitative data analysis against a set of metrics based on prior work in visual encodings, symbolic representations, and story structures.

\subsection{Activity Design}

The Data Culture Project is a free toolkit of activities for introducing various aspects of data storytelling to learners, created by co-authors Bhargava and D’Ignazio. One of the activities is called “Sketch a Story” and introduces learners to quantitative text analysis and sketching with data. Our prior work discusses the design principles \cite{bhargava_designing_2015} and pedagogy \cite{dignazio_creative_2018} behind this activity.

To summarize, the activity begins with introducing the idea of analyzing text as data and showcasing some infographic-style inspirational examples. Learners are then invited to use the WordCounter tool, a web application that generates word clouds, bigrams, and trigrams from uploaded text. Pre-loaded sample data made available to participants in iterations of this activity consisted primarily of song lyrics, with other sample data from political speeches and reports from nonprofit organizations also being available. Participants are divided into groups of two or three and given 15-20 minutes to decide which data set to use and collaboratively sketch a story based on what they find. We call these "data stories" because they build a narrative based on visual representations of data \cite{segel_narrative_2010, lee_more_2015}. During the activity we emphasize that data stories can be seeded with simple, exploratory observations, such as "Elvis Presley talks about love far more than Iron Maiden". Facilitators check-in at regular points to ensure groups are moving from data selection, to identifying a story, to physically sketching together. The materials we provide typically include crayons or markers and large pads of paper, or whiteboard markers when large whiteboards are available. The sketches they create are shared via a feedback session where they briefly introduce their story and have a lightweight response from the facilitator to connect common themes and resurface previously introduced learning goals. Further details about the activity and the supporting technology are available in the previously cited work and in the facilitator's guide on the associated website.

\subsection{Building a Corpus}

We estimate that we have run this activity more than 100 times between us, with a global footprint of learners, both in person and, more recently, online. The groups participating in this activity have included higher education students, government officials, non-profit staff. and industry professionals. After each workshop we collect the paper sketches or document them via photos. For this paper we randomly sampled sketches we have previously collected physically or digitally from across various workshops. This was a convenience sample based on which physical and digital copies were readily available and legible. We scanned and processed all of the sketches into a common digitized format.

This process left us with some gaps in our metadata about each sketch. For instance, since many of the sketches were scanned from paper and had no record of their creation data, we were unable to include temporal metadata into our analysis. Similarly, we did not capture demographic metadata about the team of authors of each sketch. We eliminated some examples from the corpus due to a lack of overall legibility or if the scan was poor quality.

Using the metrics described below as the basis for our qualitative analysis, two authors then coded a random set of 15 sketches. Following one of the conventional approaches for qualitative analysis \cite{charmaz_constructing_2006}, a single coder then coding the remainder of the corpus with particular points of uncertainty being reviewed jointly. The corpus of digital images and our qualitative coding against the metrics is available online at \href{https://doi.org/10.7910/DVN/QK35O6}{https://doi.org/10.7910/DVN/QK35O6}.

\subsection{Classification Metrics}

Based on the literature shared above, the learning goals and context of the activity, and other grounding principles in the field of data visualization, we created a set of criteria to classify the corpus of sketches. We refined this criteria through iterative analysis on a smaller set of sketches, and via discussion with colleagues in the field of data visualization education. This process resulted in a set of 12 qualitative classification metrics. Each is described below.

\subsubsection{Visual Encodings}

Inspired by Bertin’s language of visual variables, we analyzed whether a subset of key encodings were used in each sketch or not \cite{bertin_semiology_2010}.

\textbf{Position}: Is location-based grouping or other positioning used to encode some aspect of the data?

\textbf{Size}: Is the size of elements in the drawing based on data?

\textbf{Color}: Is the color of any of the elements depicted determined by the value of elements of the data?

\textbf{Shape}: Is the shape of visual elements based on data values?

\textbf{Typography}: Is the font, case, or style of any text in the sketch based on values in the data?

\subsubsection{Representations}

Our initial review revealed that most of the sketches consisted of combinations of text, symbolic representations, and graph-like depictions. In order to gauge the use of each type, we classified them separately. Each is classified on a scale from 0 (little use), 1 (some use), or 2 (extensive use). We found this 3-level scale was sufficient to capture the subjective assessment after our first assessment of inter-coder agreement - 2-level wasn't rich enough, but 5-level created significant subtle and non-meaningful disagreement. Assessment them on a 3-level scale allowed for coding sketches that used multiple types of representations and ones where a single type dominated.

\textbf{Text}: How much of the sketch is based on text-based depictions of the data?

\textbf{Symbols}: How much of the sketch is made up of symbols and iconic drawings? This could include depictions of individuals, hearts to represent love, etc.

\textbf{Graph language}: How much of the sketch is made up of visual representations classically used in graphs? This might include boxes, circles, axes, etc. 

\subsubsection{Telling a Story}

Assessing the "story" of these sketches retroactively is quite difficult. We took an approach that was based on our subjective ability as informed facilitators to understand the context and intent of each sketch’s story.

\textbf{Source}: What type of sample data did the participants decide to sketch? This categorical information captures which type of the fixed set of sample data is illustrated in the sketch - pop music lyrics, political speeches, or non-profit reports.

\textbf{Legibility}: On a scale from 0 to 2, how understandable is the story the sketch depicts? 0 indicates we are totally bewildered; 2 indicates we fully understand it, or at least think we do.

\textbf{Story Type}: What type of story is the sketch depicting? Stories are categorized as being either factoid stories (pointing out a particular stand-out point or points within the dataset), comparison stories (showing meaningful differences and similarities between parts of the data set), or change-over-time stories (documenting a change from this to that within the data set). This tiny set is by no means comprehensive, but is based on what participants very often are introduced to in other Data Culture Project activities. That taxonomy of types of stories is in turn based on our experience as data storytelling educators, work from academics \cite{segel_narrative_2010}, practitioners \cite{rosenbaum_data_2012}, and tool builders in the field \cite{tableau_best_nodate}.

\textbf{Number of Data Points}: How many data points are included in the sketch? Each depiction of metadata (ie. an artist name), word, bigram, or trigram is counted as one data point.

\subsection{A Sample}
\begin{figure}[htbp]
  \centering
  \includegraphics[width=5cm]{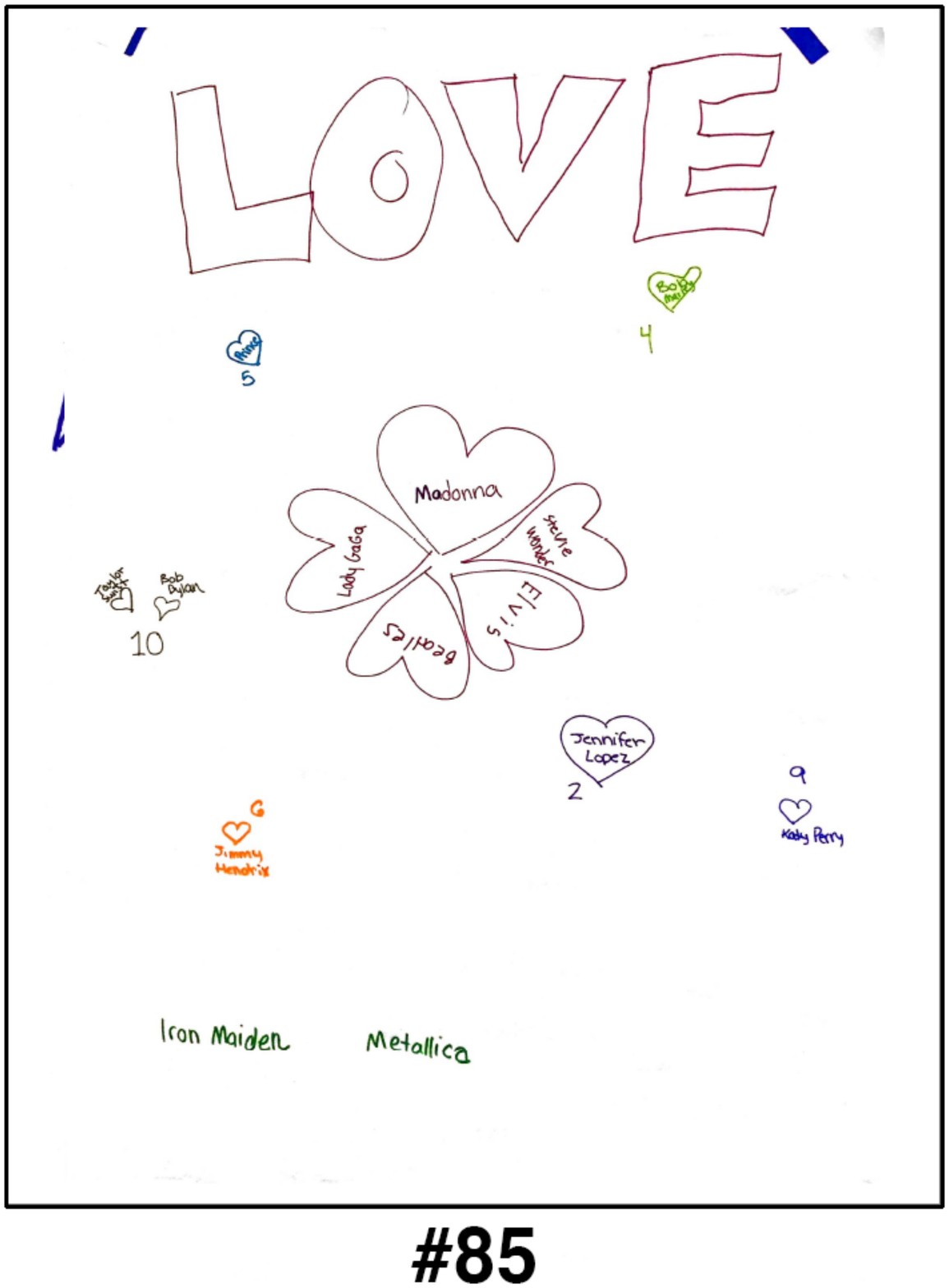}
  \caption{Sketch \#85 from our corpus.}
  \label{love}
\end{figure}

In order to illustrate these metrics in practice, below we include sketch \#85 from our corpus (Figure \ref{love}), which we've titled "Love", and discuss our coding of it. The sketch looks like a story about how often various musicians used the term "love" in their songs.

\textit{Position}: Yes. It appears that artists closer to the center of the sketch mentioned the word "love" more times.

\textit{Size}: Yes. When viewed in context with the numbers near each shape we can see that the more often an artist used "love" in their lyrics, the larger their representation was drawn.

\textit{Color}: No. While multiple colors are used, they don't appear to be based on any data.

\textit{Shape}: Yes. While the shape of the heart itself doesn't change, it's appearance next to an artist name is dictated by whether "love" was used or not.

\textit{Typography}: No. There are multiple styles of writing, but they don't appear to be based on any data.

\textit{Text}: 1 - Some use. The name of each artist, and some of the numerical quantities, make up roughly half of the elements in the sketch.

\textit{Symbols}: 1 - Some use. The 12 artists with some mention of "love" in their lyrics are represented by hearts, a very traditional symbol for depicting love.

\textit{Graph Language}: 0 - No use. There aren't any visual elements we liken to traditional graphs.

\textit{Source}: Lyrics. This sketch is based on many of the artist's lyrics included in the WordCounter tool.

\textit{Legibility}: High. It is clear what story this sketch is depicting.

\textit{Story Type}: Comparison. This sketch appears to be a comparison of the use of "love" by various artists.

\textit{Number of Data Points}: 14. This sketch depicts data from 14 different artists.

\section{Findings and Discussion}

This corpus of digitally captured and manually coded sketches of data stories reveals a number of insights into how learners approach building a practice of sketching what they see in data. Taken as a case study within the context of the activity prompt, these present insights for instructors and tool builders that build formal or informal learning experiences for novices entering the field of data visualization.

\subsection{Use of Visual Encodings}

\begin{figure}[htbp]
  \vspace*{-0.1in}
  \centering
  \includegraphics[width=7cm]{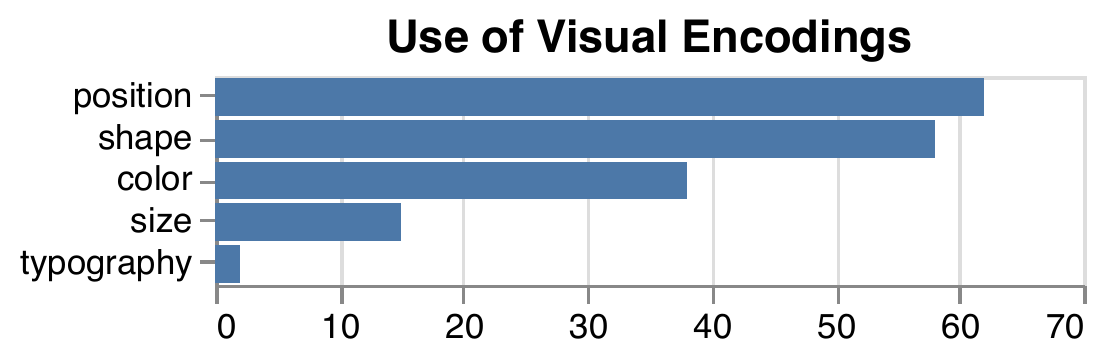}
  \caption{The number of sketches using various types of visual encodings.}
  \label{encodings}
\end{figure}

We find that position, shape, and color were the most commonly used encodings in our corpus of sketches (Figure \ref{encodings}). The prevalence of encoding data onto position and shape, both used in more than half of the sketches, suggests that novices are drawn to these channels as “standard” ways to communicate data insights. Other research suggests this learning is at least partially borne out, position being a highly effective way to communicate information about the primary quantity in a visualization \cite{kim_assessing_2018}. Size and typography, on the other hand, are encoding channels that were used very little. We find this surprising because the WordCounter tool itself encodes word frequency onto size via the word cloud it shows users. This suggests learners might need to be coaxed more to utilize these rich channels. For instance, in an educational settings an activity to introduce visual encodings might be designed to encourage learners to practice producing the same story multiple times, each built around a different encoding.  In computationally mediated contexts, one can picture how a tablet application in a visual data sketching context could be programmed to recognize a set of symbols drawn on it and be ready to suggest size scaling based on an attribute in the data. 

\subsection{Representational Variety}

\begin{figure}[htbp]
  \vspace*{0in}
  \centering
  \includegraphics[width=7cm]{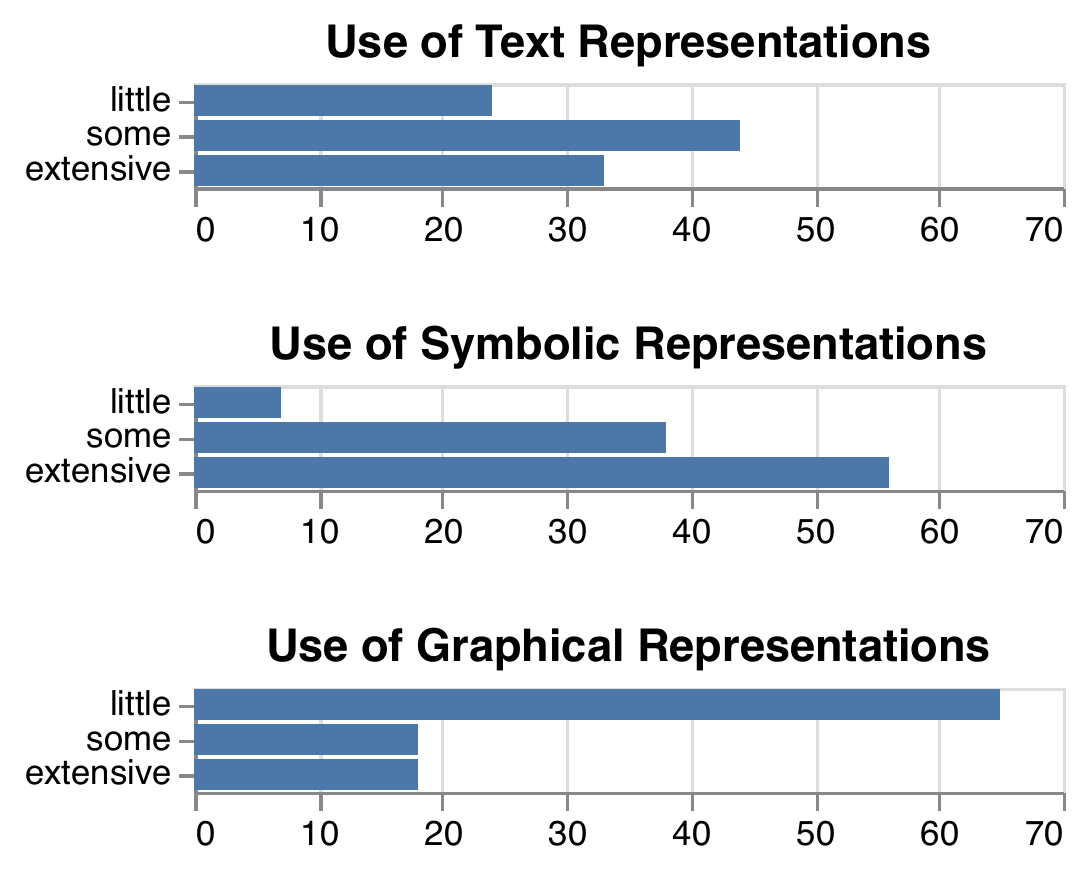}
  \caption{The use of various types of representations in our corpus.}
  \label{representations}
\end{figure}

Our activity introduces sketching very intentionally by linking it to forms participants might know already - cartoons, infographics, and more. We suspect this predisposes learners to mimic those forms, integrating combinations of textual representations. We also hypothesize that the aforementioned barrier of “design fixation” is at play here - in this short activity participants are probably hard-pressed to push beyond techniques they have seen most commonly used around them \cite{jansson_design_1991}.

At a high level there was a wide variety of balance between text, symbolic, and graphical representations (Figure \ref{representations}). Text use was mixed, with most employing it either somewhat or extensively; this is understandable since the input consisted of text and breaking away from that format may have been unintuitive for some participants. The use of symbols and graph language were the inverse of each other, with far more symbolic imagery being used than graph language. This shows a promising potential level of comfort with less pre-existing formulaic visual depictions. However, it must be kept in mind that the activity introduction typically includes instructions to “be creative” and “think outside the bar chart.” This finding on the lack of extensive use of graphical representations could certainly be a success related to one of the stated learning goals of the activity. Additionally, while graph visuals were less commonly utilized, they were still used by a sizable portion of participants despite the encouragement to use alternative formats, suggesting that breaking away from “default” ways of visualizing data can still be challenging for novices and may require more support in future endeavors.

\subsection{Story Construction}

The vast majority (77\%) of sketches were symbolic representations of musician's lyrics. The WordCounter website used in the activity provides a variety of sample datasets for use in the Sketch a Story activity, but we typically push participants towards using the music lyrics from the various artists included. This is partially because it is fun and playful data that puts them at ease and increases their willingness to take risks while learning \cite{dignazio_creative_2018}, but also because it creates more opportunities for learners to find some data they identify or have a personal relationship with.  Familiarity may breed confidence in working with data: as in Bressa, \textit{et al}, a sense of familiarity or personal connection with the dataset may have led to participants having an easier time generating stories \cite{bressa_sketching_2019}. We believe this rationale and background explains this finding.

All but 11 of the sketches were somewhat or fully legible to us as readers. For those that were somewhat legible (n=36), our memory of various workshops and our experience with the activity left us confident that we were assessing the type of story, data used, and other metrics effectively. That said, because this study is retroactive it is important to acknowledge that we are projecting intent onto the sketches that were harder to understand outside of their original context of creation.

\begin{figure}[htbp]
  \vspace*{0in}
  \centering
  \includegraphics[width=7cm]{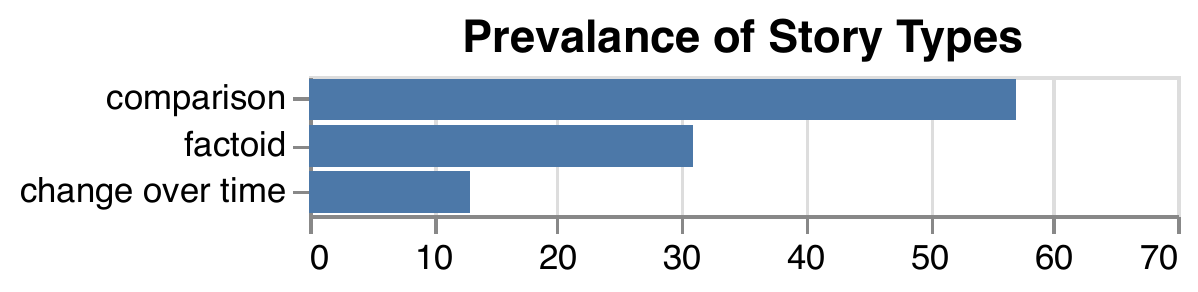}
  \caption{The number of sketches that employed each type of story.}
  \label{stories}
\end{figure}

We found that the type of story depicted varied, but the majority were comparisons between a part and a whole, or multiple parts, of some sample data (Figure \ref{stories}). This reinforces findings from Walny, \textit{et al} \cite{walny_exploratory_2015}. Novices creating the sketches in our corpus were drawn toward thinking about the data in terms of analyzing the differences and similarities between parts of the data set. This is intriguing because the WordCounter tool that supports the activity is not designed to facilitate this particular type of story. In fact users can only view the results of analyzing a single dataset at a time, i.e. you can only see multiple results if you opening two browser tabs and rearrange windows. This type of cross-artist comparison is even more difficult on mobile devices, which are being used as the main devices in this activity more and more often. Despite this higher level of friction, this shows that participants still very often employ a process of pattern-finding across datasets, rather than picking out individual items of note from one dataset. One can imagine an application of this finding with a hypothetical visualization support tool for learners that could be built to detect a new user authoring a series of comparative visualizations and in turn suggest a time-based analysis in order to broaden their design space.

Sketches tended to depict 6 or fewer data points. A full 50\% of the corpus fell in this range, with the others producing a long tail as the number of data points represented increased. This could either be an artifact of the short amount of time participants had to craft their stories (usually 15 minutes), or a sign that participants generally felt more comfortable crafting more constrained stories, rather than attempting to link together larger numbers of data points. It would take further study and observation to make more conclusive statements about the underlying cause of this finding.

\section{Conclusion}

Data storytelling isn’t simple – learning to find stories in data and present them visually is a long process that involves mastery of multiple domains. Sketching is only one piece of the puzzle, but it is one practiced by most professional designers and one that can feel daunting to newcomers. It is critical to understand how learners start sketching as part of their growth as data storytellers, and this paper offers insight into some of the patterns seen in novice sketching in a guided introductory activity across a diverse group of participants.

In this article we introduce a corpus of 101 hand-made sketches of data stories, created by a wide variety of learners in formal and informal education settings, in-person and online, over the last 10 years. We classify the sketches against a set of 12 metrics in three categories - use of visual encodings, representational variety, and story construction. We present and discuss our findings based on manually coding the corpus against these metrics. We see strong evidence that learners can easily jump into sketching data stories, setting them up well to grow into a practice of sketching elements of data representations like many experts do. In future work we hope to explore why the learners made the decisions they did, and how they describe their own drawings.

These findings contribute to our understanding of how learners begin to sketch data, and the ways in which they can be best supported on that journey. While the specific constraints and prompts of this activity certainly have impact on the sketches produced, and limit the generalizability of the findings, our analysis of this corpus contributes a strong case study. We hope it informs creators building tools for data viz, educators attempting to scaffold and extend learners, and further research into how novices adopt non techno-centric practices of data storytelling professionals. We strongly believe that studying how to introduce non-technical data practices to learners should be a central goal for data literacy researchers.

\bibliographystyle{abbrv-doi}

\bibliography{template}
\end{document}